\documentclass[
11pt]{article}
\begin{document}

\newcommand{\bra}{\langle}
\newcommand{\ket}{\rangle}
\def\tr{{\rm tr}\,}
\def\href#1#2{#2}
\newcommand{\EQ}{\begin{equation}}
\newcommand{\EN}{\end{equation}}
\newcommand{\bea}{\begin{eqnarray}}
\newcommand{\ena}{\end{eqnarray}}
\newcommand{\vs}[1]{\vspace{#1 mm}}
\newcommand{\hs}[1]{\hspace{#1 mm}}
\renewcommand{\a}{\alpha}
\renewcommand{\b}{\beta}
\renewcommand{\c}{\gamma}
\renewcommand{\d}{\delta}
\newcommand{\e}{\epsilon}
\newcommand{\s}{\sigma}
\def\bbox{{\,\lower0.9pt\vbox{\hrule \hbox{\vrule height 0.2 cm
\hskip 0.2 cm \vrule height 0.2 cm}\hrule}\,}}
\newcommand{\dsl}{\pa \kern-0.5em /}
\newcommand{\la}{\lambda}
\newcommand{\shalf}{\frac{1}{2}}
\newcommand{\pa}{\partial}
\renewcommand{\t}{\theta}
\newcommand{\tb}{{\bar \theta}}
\newcommand{\nn}{\nonumber\\}
\newcommand{\p}[1]{(\ref{#1})}
\newcommand{\lan}{\langle}
\newcommand{\ran}{\rangle}

\begin{titlepage}

\begin{center}

\hfill OU-HET 400 \\
\hfill UOSTP-01107 \\
\hfill hep-th/0112034
\vspace{2cm}

{\Large\bf Supersymmetric  D2 anti-D2 Strings}
\vspace{1.5cm}

{\large
Dongsu Bak$^a$ and  Nobuyoshi Ohta$^b$ }

\vspace{.7cm}

$^a${\it
Physics Department, University of Seoul, Seoul 130-743, Korea
\\ [.4cm]}
$^b${\it  Department of Physics, Osaka University,
Toyonaka, Osaka 560-0043, Japan\\
[.4cm]}
({\tt dsbak@mach.uos.ac.kr, ohta@phys.sci.osaka-u.ac.jp})

\end{center}
\vspace{1.5cm}

We consider the flat supersymmetric D2 and anti-D2 system, which follows
from ordinary noncommutative D2 anti-D2 branes by turning on an appropriate
worldvolume electric field describing dissolved fundamental strings.
We study the strings stretched between D2 and anti-D2 branes and show
explicitly that the would-be tachyonic states become massless.
We compute the string spectrum and clarify the induced noncommutativity
on the worldvolume. The results are compared with the matrix theory
description of the worldvolume gauge theories.
\vspace{3.5cm}
\begin{center}
\today
\end{center}
\end{titlepage}

\section{Introduction}

Recently an interesting class of 1/4 BPS brane configurations has been
constructed including the supertubes and the supersymmetric brane-antibrane
system~\cite{mateos,klee,emparan,swkim,karch}. The supertubes are tubular
configurations embedded in flat 10 dimensional space and self supported from
collapse by the contribution of angular momentum produced by their own
worldvolume gauge fields~\cite{mateos}. Many tubes separated parallelly
are again 1/4 BPS and there is no static force between them, which is shown
in the matrix theory description~\cite{klee,swkim} or the supergravity
analysis~\cite{emparan}. The tubes involve a {\it critical} electric
field\footnote{The critical value of the electric field here does not
mean the tensionless limit due to the nonvanishing magnetic components.}
as well as a magnetic field, which respectively correspond to dissolved
fundamental strings and D0 branes. The tubes have no net D2 brane charge.
Instead they carry nonvanishing  dipole components of
D2 charges~\cite{emparan,swkim}.

The elliptic deformation of tubes  discussed in Ref.~\cite{karch} still
preserves 1/4 supersymmetries and, in the limit where the ellipse becomes
two parallel lines, the system becomes a flat D2 and anti-D2 system which is
supersymmetric. (One may actually show that there are many other
supersymmetric solutions with various shapes like hyperbola for example.)
As long as the $E$-fields are critical and $B$-fields come with opposite sign,
the supersymmetries satisfied by the brane turn out to be the same as
those of the antibrane. By the matrix theory description of the worldvolume
theory, one can indeed prove that the would-be tachyons in the ordinary
D2 and anti-D2 system with $B$-fields disappear in the supersymmetric
case~\cite{karch}. Thus the system is stable against decay. In the
corresponding supergravity solutions, the D2 and anti-D2 can easily be
identified separately when they are separated. One might ask what happens
when D2 and anti-D2 are brought together and become coincident.
Clearly in this limit the dipole moment of the D2 charges becomes zero
and massless modes are expected to appear between D2 and anti-D2.
When D2 and anti-D2 are on top of each other, should we view that they
disappear to nothing? In the matrix model description, however, nothing
particular happens other than the massless degrees of freedom and the
corresponding nonabelian structure due to brane and antibrane remains.

In this note, we would like to probe the coincident limit of the
supersymmetric D2 and anti-D2 system by studying the worldsheet CFT of
strings connecting the D2 and anti-D2. Without $E$-field the tachyonic modes
of D2 anti-D2 strings are computed in Ref.~\cite{kraus} and shown to agree
with those of the matrix theory description~\cite{kraus,li,mandal}
in the zero slope limit. We begin with a generic value of $E$-field
and show how would-be tachyonic modes disappear in the string spectrum
when the $E$-field approach to the critical value. The limit we are taking
is not that straightforward; the coordinate fields $X$ become singular
in their overall factors. However the modes and their correlation functions
do not share these apparent singularity, so the worldsheet CFT is well
defined in the limit in spite of the apparent coordinate-like singularity.
We shall clarify the noncommutativity arising in the low-energy worldvolume
theories by the analysis of the commutators of the coordinate fields.
We also compute the fluctuation spectra from the matrix theory
and show that a consistent picture emerges as a consequence.

In Section 2, we study the mode expansion and spectrum of the
D2 anti-D2 strings as well as D0-D2 string spectrum. In Section 3,
we investigate the noncommutativity induced by the background gauge
field on D2 and anti-D2 and discuss the Seiberg-Witten limit.
We compare the above results to those of the matrix theory analysis
focused on the fluctuation spectra.
Last section comprises conclusions and remarks.

\section{D2 anti-D2 Strings}

We are interested in the D2-$\overline{\rm D2}$ system extended to $1,2$
directions, which is supersymmetric due to the presence of the specific
background $E$ and $B$-field~\cite{karch}. For the supersymmetric
configurations, the electric component, e.g. $E=B_{20}$ should be
critical and the magnetic part $B=B_{12}$ should come with opposite
signs on D2 and anti-D2. We begin with generic values of electric
components and take the critical limit, i.e. $E=1/(2\pi\alpha')$.
Also we take $B$ to have the same magnitude on the D2 and anti-D2.
Since $B$ comes with oppsite signatues on the D2 and anti-D2, only
D0-branes (not $\overline{\rm D0}$) are induced on them with the same
densities. The same magnitude condition of $B$-field may be relaxed as far
as they are nonvanishing on the D2 and anti-D2.

Let us examine the open string spectrum between the D2 and anti-D2.
The background field at $\s=0$ is specifically taken as
\bea
B^{(0)} = \frac{1}{2\pi \a'} \left(
\begin{array}{ccc}
0 & 0 & -e \\
0 & 0 & -b \\
e & b & 0
\end{array}
\right); \quad
g_{ij} = \eta_{ij},
\label{back0}
\ena
and $B^{(\pi)}$ at $\s=\pi$ is obtained by reversing the sign of $b$.
The boundary condition is given as
\bea
g_{ij} \pa_\sigma X^j + 2\pi \a' B_{ij} \pa_t X^j =0.
\ena
In our case of D2 at $\sigma = 0$ and anti-D2 at $\sigma = \pi$,
this is written as
\bea
\label{bc1}
\begin{array}{l}
\pa_\sigma X^0 + e \pa_t X^2 = 0, \\
\pa_\sigma X^1 \mp b \pa_t X^2 = 0, \\
\pa_\sigma X^2 + e \pa_t X^0 \pm b \pa_t X^1 = 0,
\end{array}
\quad {\rm at} \quad
\sigma=
\left\{
\begin{array}{l}
0, \\
\pi.
\end{array}
\right.
\ena
The boundary conditions for fermions are similar and the mode oscillators
can be easily obtained once the bosonic part is found.


With the specified boundary conditions, we solve the equation of motion
and find
\bea
X^0 &=& i \; \sqrt{\frac{2\a'}{1-e^2}} \sum_{n\neq 0}
 \frac{a_{n}}{n} e^{-int} \cos n\s + \frac{1}{\sqrt{1-e^2}}
(2\a' p^0 t+x^0) \nn
&&
- \; e \; \sqrt{\frac{\a'}{1-e^2}} \sum \left(
 \frac{c_{n+\nu}}{n+\nu} e^{-i(n+\nu)t - i\frac{\pi\nu}{2}}
+ \frac{d_{-n-\nu}}{n+\nu} e^{i(n+\nu)t +i\frac{\pi\nu}{2}}\right)
\sin\left[(n+\nu)\s -\frac{\pi\nu}{2}\right], \nn
\label{Xs}
X^1 &=& \sqrt{\a'} \; \sum \left(
 \frac{c_{n+\nu}}{n+\nu} e^{-i(n+\nu)t - i\frac{\pi\nu}{2}}
+ \frac{d_{-n-\nu}}{n+\nu} e^{i(n+\nu)t +i\frac{\pi\nu}{2}}\right)
\cos\left[(n+\nu)\s -\frac{\pi\nu}{2}\right] + x^1, \\
X^2 &=&
 - e \; \sqrt{\frac{2\a'}{1-e^2}}\sum_{n\neq 0} \frac{a_{n}}{n} e^{-int}
 \sin n\s
- \frac{e}{\sqrt{1-e^2}}2\a' p^0 \left( \s-{\pi\over 2}\right)  + x^2\nn
&&+\; i \; \sqrt{\frac{\a'}{1-e^2}} \sum \left(
 \frac{c_{n+\nu}}{n+\nu} e^{-i(n+\nu)t - i\frac{\pi\nu}{2}}
- \frac{d_{-n-\nu}}{n+\nu} e^{i(n+\nu)t +i\frac{\pi\nu}{2}}\right)
\cos\left[(n+\nu)\s -\frac{\pi\nu}{2}\right] , \nonumber
\ena
where $\nu$ is defined by
\bea
\tan\frac{\pi\nu}{2} = \frac{b}{\sqrt{1-e^2}}, \ \ \ \ \ \ \ \ \ \
 ( 0 \le  \nu < 1).
\ena
Here $c_{n+\nu}$'s are complex oscillator and the reality condition
of the mode expansion says that $a_n= a_{-n}^\dagger$ and
$c_{n+\nu}= d_{-n-\nu}^\dagger$.


This result shows that $\nu=1$ for $e=1$.  This implies that the level
becomes integer, which is in accordance with the supersymmetry restoration
in the limit $e=1$. Furthermore if one takes the limit $e \to 0$, Eq.~\p{Xs}
reduces to those in Ref.~\cite{kraus} for  the ordinary noncommutative
D2 and anti-D2 system which is of course unstable.
Though the $e\to 1$ limit for $X^0$ and $X^2$ looks singular in Eq.~\p{Xs},
there is no singularity in the worldsheet CFT. In fact,
the mode expansions perfectly make sense in the combinations
\bea
(\pa_\s X^0 + e \pa_t X^2) / \sqrt{1-e^2}, \quad
(e \pa_t X^0 + \pa_\s X^2) / \sqrt{1-e^2}, \nn
(e\pa_\s X^0 + \pa_t X^2) / \sqrt{1-e^2}, \quad
(\pa_t X^0 + e\pa_\s X^2) / \sqrt{1-e^2},
\ena
even in the limit. This is just like considering, in CFT, $\pa_z X$,
which has good conformal property. Alternatively, one may explicitly compute
the correlation functions $\bra X^i (z)X^j(z') \ket$ and verify
that they are nonsingular in the limit $e \rightarrow 1$.

Upon quantization, we posit the nonvanishing commutation relations
between oscillators to be
\bea
&& [a_m,a_n]= -\, m \delta_{m+n} ,(m\neq 0)\,; \ \ \ \  [x^0,p^0]=-i;
\nn
&& [c_{m+\nu}, d_{n-\nu}]=({m+\nu}) \delta_{m+n}\,.
\label{commutators}
\ena
The justification of the quantization here is relegated to the appendix.
One thing to note is that we have not yet specified the
commutation relation between $x^i$'s and these will be fixed
in the next section by requiring the equal-time commutators of $X^i(\s)$
to commute with each other for $0< \s <\pi$.
The corresponding Virasoro generators can be found as
\bea
L_m={1\over 2} \sum_n\Bigl(-a_{m-n} a_n
+ c_{m-n+\nu} d_{n-\nu} + d_{n-\nu} c_{m-n+\nu}\Bigr)\,.
\ena
with  $L_0$ being the Hamiltonian of the system.

We define vacuum by $ c_{n+\nu}|0\rangle =0$ for $n \ge 0$
and $d_{n-\nu} |0\rangle =0 $ for $n >0$.
Hence the corresponding vacuum energy from $L_0$ becomes
\bea
E_\nu = -{1\over 2}\sum^\infty_{n=1} n
+{1\over 2}\sum^\infty_{n=1} (n-\nu) +
{1\over 2}\sum^\infty_{n=1} (n-(1-\nu))
= +{1\over 24}+ \left( {1\over 24}-{1\over 8}(2\nu-1)^2\right)\,.
\ena
Note that this is the contribution to the vacuum energy from $X^0, X^1, X^2$.

The vacuum energy for the Ramond sector is trivial; the bosonic one is
canceled by the fermionic contribution and the total is zero.
In the Neveu-Schwarz sector, the vacuum energy from the NS-fermions
associated with $0,1,2$ components is given by the substitution $\nu \to
|\nu-1/2|$. In the ``light-cone'' gauge, the contributions from $a_n$ and
one transverse oscillator cancel with each other. Summing the contributions
of the rest of oscillators, we find the ground state energy
\bea
E_\nu^{total} &=& \left( {1\over 24}-{1\over 8}(2\nu-1)^2\right)
 - \left( {1\over 24}-{1\over 8}(2|\nu-1/2|-1)^2\right)
-\frac{6}{24} - \frac{6}{48} \nn
&=&-{1\over 4} -\frac{|\nu -1/2|}{2} \,.
\label{g}
\ena
For $\nu=0$, this is consistent with the vacuum energy of
NS sector without background field.

The ground state energy~\p{g} gives two lower states with energies
\bea
E_1 = -\frac{1}{2} \nu, \;\; &{\rm for}& \;\; 1 \geq \nu \geq \frac{1}{2}, \nn
E_0 = \frac{1}{2} (\nu-1) , \;\; &{\rm for}& \;\; \frac{1}{2} \geq \nu \geq 0.
\ena
When $\nu=0$, $|E_0\ran$ gives the true ground state and $|E_1\ran$ is
the first excited state, but the energy changes when $\nu$ is increased.
For $\nu \geq \frac{1}{2}$, $|E_1\ran$ becomes the true ground state and
$|E_0\ran$ is the first excited state.
For $\nu=0$ and D2-D2, the ground state $|E_0\ran$ is projected out by
GSO projection and $|E_1\ran$ is kept. However, our D2-$\overline{\rm D2}$
system has opposite GSO projection, and the state with $E_1$ is projected
out and $|E_0\ran$ is kept, giving tachyonic state. This state becomes
massless for $\nu=1$, giving a stable system.

Let us consider what would be the spectrum at lower levels. The ground state
is denoted by
$\left| -\nu/2\,\right\ran $ .
Corresponding to the mode oscillators $c_{n+\nu}$ and $d_{-n-\nu}$,
we have fermionic oscillators $\psi_{n+\nu-\frac12}$ and
$\bar\psi_{-n-\nu-\frac12}$. For $\nu \geq \frac12$,
$\bar\psi_{-\nu+\frac12}, \psi_{\nu-\frac32}$ and transverse oscillators
$\psi^s_{-\frac12} (s=3,\ldots,8)$ give lower states (which remain after
GSO projection)
\bea
\left| (\nu\!-\!1)/2\, \right\ran \ \equiv
\bar \psi_{-\nu+\frac12} \left| -\nu/2\, \right\ran ,\quad
\left| 3(1\!-\!\nu)/2 \,\right\ran \equiv
\psi_{\nu-\frac32} \left| -\nu/2\, \right\ran ,\quad
\left| (1\!-\!\nu)/2\, \right\ran^s \equiv
\psi^s_{-\frac12} \left| -\nu/2\, \right\ran .
\ena
For $\nu<1$, the first state is the ground state discussed above and
gives tachyonic one. All these states give 8 massless for $\nu=1$,
and the level is degenerate with Ramond sector, in accordance with the
restoration of supersymmetry. In Ref.~\cite{karch}, it was argued that
1/4 supersymmetry is restored. Our result of 8 massless bosonic states is
consistent with this claim because the representation of 8 supercharges
contains $2^4$ states in total, half of which are bosonic.

It is interesting that the limit $\nu=1$ can be achieved just by sending
$e \to 1$ but $b$ finite. In the absence of electric background,
$b$ must be sent to infinity in order to achieve supersymmetry. However,
that is incompatible with the Seiberg-Witten limit to be discussed in
the next section.

The physical picture of the system can be most easily understood by
combining T-duality~\cite{karch,cho,OT} and Lorentz boost. Let us make
T-duality in the $X^2$-direction,  and then make the Lorentz boost in
the same direction by
\bea
\left(\begin{array}{c}
X^0{}' \\
X^2{}'
\end{array} \right)
=
\left(\begin{array}{cc}
\cosh \b & \sinh \b \\
\sinh \b & \cosh \b
\end{array} \right)
\left(\begin{array}{c}
X^0 \\
X^2
\end{array} \right); \quad
\cosh \b = \frac{1}{\sqrt{1-e^2}},\;\;
\sinh \b = \frac{e}{\sqrt{1-e^2}}.
\ena
The boundary conditions \p{bc1} are then cast into
\bea
\begin{array}{l}
\pa_\s X^0{}' = 0, \\
\label{bbc1}
\pa_\s ( \sqrt{1-e^2} X^1 \mp b X^2{}' ) = 0, \\
\pa_t ( b X^1 \pm \sqrt{1-e^2} X^2{}' ) = 0,
\end{array}
\quad {\rm at} \quad
\s=
\left\{
\begin{array}{l}
0, \\
\pi.
\end{array}
\right.
\ena
This means that the system consists of two D1-branes, one tilted in the
$(X^1,X^2{}')$ plane by the angle $\frac{\pi}{2}\nu$ at $\s=0$ and
another by $-\frac{\pi}{2}\nu$ at $\s=\pi$.
There is no supersymmetry in the tilted two D1-branes. In the limit $\nu \to 1$,
however, this system reduces to boosted parallel branes, and supersymmetry
is restored. The number of restored supersymmetry cannot be determined by
simply looking at these boundary conditions, as is usual for rotated brane
configuration. It was determined 1/4 by Killing spinor analysis~\cite{karch}.


We would like to comment upon the 0-2 string spectra at this point.
We consider D0 at $\sigma = 0$ and D2 at $\sigma = \pi$, for which
the boundary conditions become
\bea
\left\{
\begin{array}{l}
\pa_\sigma X^0  = 0, \\
\label{bc10}
\pa_t X^2 = 0, \\
\pa_t X^1 = 0,
\end{array}\right.
\quad {\rm at} \;\; \sigma=0; \qquad
\left\{
\begin{array}{l}
\pa_\sigma X^0 + e \pa_t X^2 = 0, \\
\pa_\sigma X^1 - b \pa_t X^2 = 0, \\
\pa_\sigma X^2 + e \pa_t X^0 +b \pa_t X^1 = 0,
\end{array}\right.
\quad {\rm at} \;\; \sigma=\pi.
\ena
For this case, we find
\bea
X^0 &=& i \; \sqrt{\frac{2\a'}{(1-e^2)}} \sum_{n\neq 0}
 \frac{a_{n}}{n} e^{-int} \cos n\s + \frac{1}{\sqrt{1-e^2}}
(2\a' p^0 t+x^0) \nn
&&
+ \; e \; \sqrt{\frac{\a'}{1-e^2}} \sum \left(
 \frac{c_{n+\nu'}}{n+\nu'} e^{-i(n+\nu')t}
+ \frac{d_{-n-\nu'}}{n+\nu'} e^{i(n+\nu')t }\right)
\cos(n+\nu')\s, \nn
\label{Xs0}
X^1 &=& \sqrt{\a'} \; \sum \left(
 \frac{c_{n+\nu'}}{n+\nu'} e^{-i(n+\nu')t }
+ \frac{d_{-n-\nu'}}{n+\nu'} e^{i(n+\nu')t }\right)
\sin (n+\nu')\s + x^1, \\
X^2 &=&
 - e \; \sqrt{\frac{2\a'}{1-e^2}}\sum_{n\neq 0} \frac{a_{n}}{n} e^{-int}
 \sin n\s
- \frac{e}{\sqrt{1-e^2}}2\a' p^0 \left( \s-{\pi\over 2}\right)  + x^2\nn
&&+\; i \; \sqrt{\frac{\a'}{1-e^2}} \sum \left(
 \frac{c_{n+\nu'}}{n+\nu'} e^{-i(n+\nu')t}
- \frac{d_{-n-\nu'}}{n+\nu'} e^{i(n+\nu')t}\right)
\sin (n+\nu')\s , \nonumber
\ena
where $\nu'$ is defined by
\bea
e^{2i\pi\nu'} = -{\sqrt{1-e^2}+ib\over
\sqrt{1-e^2}-ib} = e^{i\pi (\nu+1)}, \ \ \ \ \ \ \ \ \ \
 ({1}/{2} \le  \nu' < 1).
\label{newnu}
\ena
Then the expressions for the commutation relation between oscillators,
the Hamiltonian, the ground state, the excited spectrum take the same forms
as before. So the ground states after GSO projection give 8 massless
bosonic degrees again but with the newly defined $\nu'$.

\section{Noncommutativity and Seiberg-Witten Limit}

We shall compute here the noncommutativity induced due to the
background gauge fields of the D2 anti-D2 system and investigate the
Seiberg-Witten decoupling limit involved with the system.

For this let us first compute the equal-time commutators of $X^i(\s)$.
Using the commutation relations in (\ref{commutators}), we have
\bea
[X^1(\s), X^2(\s')]= [x^1,x^2]
-{2i \a'\over \sqrt{1-e^2}}\sum_n{1\over n+\nu}
\cos\left((n\!+\!\nu)\s \!-\!\frac{\pi\nu}{2}\right)
\cos\left((n\!+\!\nu)\s'\! -\!\frac{\pi\nu}{2}\right)
\ena
To evaluate this, we use the identities
\bea
&&\sum^\infty_{n=-\infty} {\cos n\theta\over n+a} =
{\pi \over \sin a\pi}\cos\left(a(2m+1)\pi -a\theta \right)
\ \ \ \ \  {\rm for} \ \  2m\pi \le\theta \le 2(m+1)\pi\nn
&&\sum^\infty_{n=-\infty} {\sin n\theta\over n+a} =\left\{
\begin{array}{cl}
{\pi \over \sin a\pi}\sin\left(a(2m+1)\pi -a\theta \right)
&  \mbox{for $2m\pi <\theta < 2(m+1)\pi$} \\
 0 & \mbox{for $\theta= m\pi$}
\end{array}
\right.
\label{iden}
\ena
with an integer $m$ and $a\neq 0$.
One then finds that
\bea
\sum_n {1 \over n+\nu}
{\cos\left((n\!+\!\nu)\s \!-\!\frac{\pi\nu}{2}\right)
\cos\left((n\!+\!\nu)\s'\! -\!\frac{\pi\nu}{2}\right)}
=\pi
\cot {\nu\pi\over 2}
\left\{
\begin{array}{cl}
1
&  \mbox{if $0 \!<\! \s\!+\!\s' \!<\! 2\pi$} \\
 \cos \nu\pi & \mbox{if $\s=\s'=0,\ \pi$}
\end{array}
\right.
\ena
By requiring the bulk contribution to be zero, we fix
the commutator $[x^1,x^2]$ as
\bea
[x^1,x^2]=
{i \a'\pi\over b}\,.
\ena
The commutation relation becomes
\bea
[X^1(\s), X^2(\s')]|_{\s=\s'=0,\pi}=i {2\pi\a' b\over 1+b^2-e^2}\,,
\ena
while vanishing in the bulk.

For the $0-2$ commutator, we get
\bea
&&[X^0(\s), X^2(\s')]= \frac{1}{\sqrt{e^2-1}}[x^0,x^2]
+i{2\a' e\over 1-e^2}(\s'-\pi/2) +i{2\a' e\over 1-e^2}\sum_{n\neq 0}{1\over n}
\cos n\s \sin n\s'
\nn
&&\ \ \ \ \ \ \ \ \ \ \ \ \ \ \ \ \
+i{2\a' e \over 1-e^2}\sum_n{1\over n+\nu}
\sin\left((n\!+\!\nu)\s \!-\!\frac{\pi\nu}{2}\right)
\cos\left((n\!+\!\nu)\s'\! -\!\frac{\pi\nu}{2}\right)\,.
\ena
In addition to (\ref{iden}), we shall use also the identity
\bea
\sum_{n\neq 0} {\sin n\theta\over n}= \left\{
\begin{array}{cl}
 \pi-\theta  &  \mbox{for $0 <\theta < 2\pi$}\\
 -\pi-\theta  &  \mbox{for $-2\pi <\theta < 0$} \\
 0 & \mbox{for $\theta= 0$}
\end{array}
\right.
\ena
The straightforward evaluation leads to
\bea
[X^0(\s), X^2(\s')]|_{\s=\s'=0}=
-[X^0(\s), X^2(\s')]|_{\s=\s'=\pi}
= -i {2\pi\a' e\over 1+b^2-e^2}\,,
\ena
where $[x^0,x^2]=0$ is chosen such that there is again no bulk contribution
to the commutator. Finally, with $[x^0,x^1]=0$,
$[X^0(\s), X^1(\s')]=0$
including boundaries.  Hence we see that the end points of the
string become noncommutative. In particular, the contribution
of $[X^1(\pi),X^2(\pi)]$ has the opposite signature to the one given in
Ref.~\cite{chu}. This is because they consider strings ending on
D-branes with the same $b$ while we are considering here
strings from D2  with $b$ ending on anti-D2 with $-b$.

The result can be neatly written as
\bea
[ X^i(\s), X^j(\s')] = i\Theta^{ij,(\s)}\e_{\s,\s'},
\ena
where $\Theta^{ij,(\s)}$ is the Seiberg-Witten expression of noncommutativity
\bea
\Theta^{ij,(\s)}=2\pi\a'\left(\frac{1}{g+2\pi\a'B^{(\s)}}\right)^{ij}_A
=-(2\pi\a')^2
\left(\frac{1}{g+2\pi\a'B^{(\s)}}B^{(\s)}\frac{1}{g-2\pi\a'B^{(\s)}}
\right)^{ij},
\ena
where the background $B^{(\s)}$ is specified in Eq.~\p{back0} and
is nonvanishing only at $\s=0,\pi$, and $\e_{\s,\s'}$ is defined to be
$\pm 1$ for $\s=0,\pi$.

To discuss the Seiberg-Witten decoupling limit,
we restore $g_{ij}={\rm diag}(- |g_{00}|, g_{11},g_{22})$ and take the limit
where $g_{11},\ g_{22}\sim \e$ and $\a'\sim \sqrt{\e}$ while fixing
$B_{02}$, $B_{12}$ and $g_{00}$~\cite{karch}.
The critical condition becomes $e^2=|g_{00}|g_{22}$, for which the
system is supersymmetric. The noncommutativity at the end points of
the open string becomes
\bea
&&[ X^1(0), X^2(0)]=
[ X^1(\pi), X^2(\pi)]=
i {2\pi\a' b\over g_{11}g_{22}\left(1-{e^2\over |g_{00}|g_{22}}\right)+
b^2}\,,\nn
&&[ X^0(0), X^2(0)]=
-[ X^0(\pi), X^2(\pi)]=
-i {2\pi\a' e\over |g_{00}|g_{22}\left(1-{e^2\over |g_{00}|g_{22}}+
{b^2\over g_{11}g_{22}}\right)}\,.
\ena
Now using the critical condition and taking the limit in which
$g_{11},\ g_{22}\sim \e$ and $\a'\sim \sqrt{\e}$ while fixing
$B_{02}$, $B_{12}$ and $g_{00}$, we obtain
\bea
[ X^1(0), X^2(0)]=
\ [ X^1(\pi), X^2(\pi)]={i\over B}\,,
\label{noncommutativity}
\ena
with $B=b/(2\pi \a')$ and all other coordinate commutators vanish
especially at the end points.

Thus we end up only with the spatial noncommutativity in the Seiberg-Witten
limit~\cite{karch}. This implies that the worldvolume theory in the zero slope
limit may be described by the noncommutative gauge theory with spatial
noncommutativity $[x,y]=i\theta$. In (\ref{noncommutativity}), it
looks like that the noncommutativities on both branes have the same
signature. But, taking into account of the fact that the two ends of open
strings are oppositely charged, the noncommutativity geometry on the D2 is
with $\theta=1/B$ while on the anti-D2 with $\theta=-1/B$. This suggests
the the effective low-energy theory on this system is a noncommutative
super Yang-Mills theory from D2-branes and that with opposite noncommutativity
from anti-D2-branes, together with bifundamental scalars from the open
strings between them. For the field theory description of the worldvolume
theory, the use of different noncommutativities on D2 and anti-D2 is
inconvenient and not conventional. Instead, one may use the description
where $\theta=1/B$ for both D2 and anti-D2 and the effects of opposite
noncommutativity on the anti-D2 is described by the background magnetic field.
Here the background magnetic field is not decoupled in general, but
 the $U(2)$ noncommutative
gauge symmetries are still in effect.
As shown in Refs.~\cite{kraus,li,mandal}, the background magnetic field
then makes the spectrum tachyonic for the ordinary noncommutative
D2 and anti-D2. In our case, the would-be tachyonic degrees disappear,
in spite of the presence of the same background magnetic field,
and the massless continuum spectra  appear as the above analysis of
the string spectrum indicates. To study the details, one has to look at
the interaction amplitudes obtained from the worldsheet, which goes beyond
the scope of this short note. The worldvolume gauge theory has been obtained
in the matrix theory description~\cite{karch} and we shall use this for
further check of the continuum spectra of the fluctuation.

\section{Comparison to the Matrix Theory Description}

In the matrix theory description~\cite{karch}, it was shown that
the supersymmetric D2 anti-D2 system is described, in the gauge $A_0=Y$,
by the background\footnote{Unlike the notation in \cite{karch},
we use here $Y$ as the second worldvolume matrix
coordinate and $Z$ as a transverse direction.}
\begin{eqnarray}
X=\left(
\begin{array}{cc}
x & 0\\
0 & x
\end{array}
\right)\,,\ \ \ Y=\left(
\begin{array}{cc}
y & 0\\
0 & -y
\end{array}
\right)
\end{eqnarray}
with
\begin{eqnarray}
x+iy=   
\sqrt{2\theta}\sum^\infty_{n=0}\sqrt{n+1} |\,n\,\rangle\langle n+1|\,.
\end{eqnarray}
This satisfies $[x,y]=i\theta$ with the noncommutativity scale $\theta$
being  $2\pi \a'/b$ in the previous sections. The dynamics of strings
connecting D2 and anti-D2 is described by the off-diagonal fluctuations of
the above matrices. Hence we shall turn on these off diagonal components by
\begin{eqnarray}
X=\left(
\begin{array}{cc}
x & H_1\\
H^\dagger_1 & x
\end{array}
\right)\,,\ \ \ Y=\left(
\begin{array}{cc}
y &  H_2\\
 H^\dagger_2 & -y
\end{array}
\right)\,,\ \ \ X_s=\left(
\begin{array}{cc}
0 & H_s\\
H^\dagger_s & 0
\end{array}
\right)\,,
\end{eqnarray}
where the index $s=3,4,\cdots,9$ refers to the transverse matrix
coordinates including $Z=X_3$.
The linear fluctuations are governed by the equations
of motion
\begin{eqnarray}
D_0D_0 X_J+[X_I,[X_I, X_J]]=0
\end{eqnarray}
together with the Gauss law
$[X_I,D_0X_I]=0$.
In the gauge $A_0=Y$ again, we find that the Gauss law leads to
\begin{eqnarray}
\theta \partial_y \dot{H}_1 -2iy \dot{H}_2
-\theta^2 \partial^2_y H_2 - 4i\theta H_1 -
2i\theta y\partial_y H_1 =0\,,
\label{gauss}
\end{eqnarray}
where we have used $[x,\ ]=i\theta \partial_y$ and transformed all the
matrix variables to ordinary functions using Weyl-Moyal mapping.
Consequently, the variables $H$'s appearing in the above equation
are now ordinary functions with all the products here ordinary.
Similarly from the other components, one obtains
\begin{eqnarray}
&& \ddot{H}_1 -4iy \dot{H}_1 -\theta \partial_y \dot{H}_2
=0 \,,\nn
&& \ddot{H}_2 -2iy \dot{H}_2
-\theta^2 \partial^2_y H_2 - 4i\theta H_1 -
2i\theta y\partial_y H_1 =0\,,\nn
&& \ddot{H}_s -4iy \dot{H}_s -\theta^2 \partial^2_y {H}_s
=0\,.
\label{fluctuation}
\end{eqnarray}
Combining (\ref{gauss}) and (\ref{fluctuation}), one finds that $H_2$
is related to $H_1$ by
\begin{eqnarray}
 \dot{H}_2= \theta\partial_y H_1\,,
\end{eqnarray}
while $H_1$ and $H_s$ satisfies
\begin{eqnarray}
\ddot{H} -4iy \dot{H} -\theta^2 \partial^2_y {H}
=0\,.
\label{fluctuation1}
\end{eqnarray}
Hence there are eight independent degrees are present
in the fluctuations. For the case of the harmonic time
dependence $H= h(x,y) e^{-i\omega t}$, the equation becomes
\begin{eqnarray}
-\partial^2_y h +{4 \omega\over \theta^2} \left(y -{\omega\over 4}\right) h
=0  \,.
\label{airy}
\end{eqnarray}
which corresponds to the Airy equation.
The matrix Hamiltonian is proportional to $\omega^2 H^2$
and clearly nonnegative definite for the fluctuation.

Since the  spectrum is continuous, we conclude that the degrees involved
are massless. However the equation is not a free wave equation but involves
a peculiar background proportional to $y$ that is not enough to make
the spectrum discrete. Hence the results here are consistent with
the string theory analysis especially in the number of the
massless degrees. Of course, there are no tachyons in the spectrum
as found from the string spectrum.  Finally, we note that
a multiplication of an arbitrary function of $x$ only to $h$
still solves the equation without affecting $\omega$. This
property will disappear if one considers nonlinear
corrections to the equations of motion.

\section{Conclusions}

In this note, we have considered the strings connecting D2 and anti-D2 that
are supersymmetric due to the background $E$ and $B$ fields. Beginning with
generic values of $E$-field, we have shown that the tachyons disappear in
the critical limit corresponding to the supersymmetric configurations.
Although there appear apparent singularities in the coordinate fields,
the worldsheet CFT is well defined. In the limit, the ground states
after GSO projection become massless, which is consistent with the
restoration of the supersymmetries. For the low-energy description
implied by the worldsheet CFT, one has to compute the three and four
point amplitudes to extract the interaction terms in the zero slope limit.
Instead of going this direction, we investigate the matrix theory fluctuation
spectra corresponding to the D2 and anti-D2 strings and found that the
massless degrees indeed appear in spite of the presence of the background
fields. Consequently a coherent picture from both descriptions emerges.

The original supertube has a circular geometry and the elliptic
deformation preserves again a quarter of the supersymmetries. Probing
these tubes by strings are of interest though the analysis are expected
to be more involved than those presented here. One may ask how the radial
size and the background $E$ and $B$ fields are related to the disappearance
of tachyonic modes or how the 0-tube strings behave. These kinds of
information ought to be helpful in resolving the dynamical issues
involved with the appearance of tubes out of F1 and D0's or what
governs the shapes~\cite{swkim,bena}.

The dynamics  we are ultimately interested in are the dynamical processes
in which  an initial collection of D0's and F1 flows into the tubular
branes or the supersymmetric D2 and anti-D2 systems or vice versa.
Such deformations are large in the sense that we call only local
bounded-energy fluctuations small. Along these large deformation including
the change of the radius, $E$ and $B$-field, it is particularly of interest
to know how the stability of the systems are affected.

\noindent{\large\bf Acknowledgment}

We would like to thank Jin-Ho Cho, Pillial Oh,
Hyeonjoon Shin
and  Piljin Yi  for discussions.
The work of DB was supported in part by KOSEF 1998
Interdisciplinary Research Grant 98-07-02-07-01-5.
That of NO was supported in part
by a Grant-in-Aid for Scientific Research No. 12640270, and by a
Grant-in-Aid on the Priority Area: Supersymmetry and Unified Theory of
Elementary Particles.

\appendix
\section{Quantization of the System}
\setcounter{equation}{0}
\renewcommand{\theequation}{A.\arabic{equation}}
We justify  the
quantization  appearing in (\ref{commutators}) by
the following reverse procedure. Note first that we do not yet
specify our system by a specified action. We define our
system  by the following
first order form
\bea
S= \int dt\left(-{i\over 2}\sum_{n\neq 0}{1\over n} \dot{a}_n a_{-n}
-\dot{x}^0 p^0
+{i}\sum_{n}{1\over n+\nu} \dot{c}_{n+\nu} d_{-n-\nu}
- L_0
)\right)\,,
\label{action}
\ena
where arbitrary time dependence of the oscillator variables are to be
understood. Upon quantization, the equal time commutation relations in
(\ref{commutators}) follows. In connection with the original system,
we define $X^i$ such that the oscillator variables in (\ref{action})
replaces the corresponding oscillator variables absorbing the time
dependent phase factors of (\ref{Xs}). Then the Heisenberg equation of
motion of $X^i$ implies that $X^i$ satisfies the desired free wave equation.
Furthermore the time development of $X^i$  is consistent with the boundary
conditions in (\ref{bc1}) on shell. If one turns the action in (\ref{action})
into second order form and rewrite oscillator variables in terms of $X^i$,
one then can show in fact that the second order action is proportional to
${T\over 2}\int dt d\s (\dot{X}\cdot\dot{X}-
\partial_\s X\cdot \partial_\s X)$ up to boundary terms that may be
relevant in specifying the boundary conditions. This completes the
definition of our system of the D2 anti-D2 strings.

Another way to derive the commutation relations of the oscillators is
to use the ``canonical'' momenta defined by
\bea
P_i = \frac{1}{2\pi\a'} (g_{ij} \pa_t X^j + 2\pi\a' B_{ij}^{(\s)}\pa_\s X^j),
\ena
and impose the equal-time commutation relations
\bea
[X^i(\s,t), P_j(\s',t) ] = i \delta^i{}_j\d(\s-\s'),
\label{etcr}
\ena
with the delta function specified by the Neumann boundary
conditions.  Or again reversing the procedure, the momenta defined
above satisfy the equal time commutation relations (\ref{etcr})
provided the commutation relations of the oscillator variables
in (\ref{commutators}) holds.

In this procedure, care must be
taken of the fact that the background changes
depending on $\s$, as specified in Eq.\p{back0}. In fact as far as
the boundary value of $B_{ij}(\s)$ is specified as in Eq.~\p{back0},
how to extend it inside bulk does not matter. One may prove this fact
by the explicit computation using the commutation relation.

For the sake of illustration, we here present a proof only for
the case $i=j=1$. First note that
\bea
[X^1(\s), \dot{X}^1(\s') ] = 2 i \a'
\sum_n \cos\left[(n+\nu)\s -\frac{\pi\nu}{2}\right]
\cos\left[(n+\nu)\s' -\frac{\pi\nu}{2}\right]\,,
\ena
where we have used the commutation relations.
Using $\sum_n\sin n\theta =0$, one may rearrange the above
as
\bea
[X^1(\s), \dot{X}^1(\s') ] =  i \alpha'
\sum_n \left[\,\cos n(\s\!+\!\s')\cos \nu(\s\!+\!\s'\!-\!\pi)
+ \cos n(\s\!-\!\s')\cos\nu(\s\!-\!\s')\,\right]\,.
\ena
Now we use the fact that
\bea
\sum_n\cos n x = 2\pi \sum_k \tilde\delta ({x-2k\pi})
\ena
where $\tilde\delta (x)$ is the usual delta function defined as
$\int^{\infty}_{-\infty} \tilde\delta ({x- x_0}) f(x)= f(x_0)$ for any
continuous function $f$. One then obtains
\bea
[X^1(\s), \dot{X}^1(\s') ] = 2\pi \alpha' i
\left[\,
\tilde\delta (\s-\s') +\left( \tilde\delta (\s+\s')
+\tilde\delta (\s+\s'-2\pi)\right)\cos \nu\pi
\,\right]
\ena
for $0\le\s,\s'\le \pi$. By a similar computation, one has
\bea
[X^1(\s), \partial_{\s'}{X}^2(\s')] = {4\pi \alpha' i b\over
1+b^2-e^2}\left(- \tilde\delta (\s+\s')
+\tilde\delta (\s+\s'-2\pi)\right)
\ena
again for $0\le\s,\s'\le \pi$. Combining these two results,
one finally gets
\bea
[X^1(\s), P_1(\s') ] =i\left[
\tilde\delta (\s-\s') + \tilde\delta (\s+\s')
+\tilde\delta (\s+\s'-2\pi)\right] = i\delta (\s-\s')
\ena
for $0\le\s,\s'\le \pi$. This completes the proof for $i=j=1$
case and the remaining can be proved in a similar way.

Finally using again (\ref{commutators}) and regulating
appropriately, one can show that the momenta commutes with each other
at equal time, i.e. $[P_i(\s,t), P_j(\s',t) ]=0$.



\begin{thebibliography}{99}

\bibitem{mateos}
D.~Mateos and P.~K.~Townsend,
Phys.\ Rev.\ Lett.\  {\bf 87} (2001) 011602, hep-th/0103030.

\bibitem{klee}
D.~Bak and K.~Lee,
Phys.\ Lett.\ B {\bf 509} (2001) 168, hep-th/0103148.

\bibitem{emparan}
R.~Emparan, D.~Mateos and P.~K.~Townsend,
JHEP {\bf 0107} (2001) 011, hep-th/0106012.

\bibitem{swkim}
D.~Bak and S.~W.~Kim,
hep-th/0108207.

\bibitem{karch}
D.~Bak and A.~Karch,
hep-th/0110039.

\bibitem{kraus}
P.~Kraus, A.~Rajaraman and S.~H.~Shenker,
Nucl.\ Phys.\ B {\bf 598} (2001) 169, hep-th/0010016.

\bibitem{li}
M.~Li,
Nucl.\ Phys.\ B {\bf 602} (2001) 201, hep-th/0010058.

\bibitem{mandal}
G.~Mandal and S.~R.~Wadia,
Nucl.\ Phys.\ B {\bf 599} (2001) 137, hep-th/0011094.

\bibitem{cho}
J.~H.~Cho and P.~Oh,
Phys.\ Rev.\ D {\bf 64} (2001) 106010, hep-th/0105095.

\bibitem{OT}
N.~Ohta and D.~Tomino,
Prog.\ Theor.\ Phys.\  {\bf 105} (2001) 287, hep-th/0009021.

\bibitem{chu}
C.~S.~Chu and P.~M.~Ho,
Nucl.\ Phys.\ B {\bf 550} (1999) 151, hep-th/9812219.

\bibitem{bena}
I.~Bena,
hep-th/0111156.

\end{thebibliography}
\end{document}